\documentclass[twocolumn,prd,floatfix,preprintnumbers,a4paper,nofootinbib,superscriptaddress]{revtex4-1}
\usepackage{epsfig}
\usepackage{graphics}
\usepackage{graphicx}
\usepackage{amsmath,amssymb}
\usepackage{amsfonts}
\usepackage{bm}
\usepackage[usenames,dvipsnames]{xcolor}
\usepackage{amssymb}
\usepackage{psfrag}
\usepackage{times}
\usepackage[varg]{txfonts}
\usepackage{xspace}
\usepackage{float}
\usepackage{placeins} 
\usepackage{capt-of} 

\usepackage[colorlinks, pdfborder={0 0 0}]{hyperref} 
\definecolor{LinkColor}{rgb}{0.75, 0, 0}
\definecolor{CiteColor}{rgb}{0, 0.5, 0.5}
\definecolor{UrlColor}{rgb}{0, 0, 0.75}
\hypersetup{linkcolor=LinkColor}
\hypersetup{citecolor=CiteColor}
\hypersetup{urlcolor=UrlColor}

\let\oldtheequation\theequation
\makeatletter
\def\tagform@#1{\maketag@@@{\ignorespaces#1\unskip\@@italiccorr}}
\renewcommand{\theequation}{(\oldtheequation)}
\makeatother

\usepackage[utf8]{inputenc}
\usepackage{ulem}
\newcommand{\bns}[1][]{binary neutron star#1 (BNS#1)\renewcommand{\bns}[1][]{BNS##1\xspace}\xspace}
\newcommand{\nss}[1][]{Neutron stars#1 (NSs#1)\renewcommand{\nss}[1][]{NSs##1\xspace}\xspace}
\newcommand{\ns}[1][]{Neutron star#1 (NS#1)\renewcommand{\ns}[1][]{NS##1\xspace}\xspace}
\newcommand{\eos}[1][]{equation of state#1 (EOS#1)\renewcommand{\eos}[1][]{EOS##1\xspace}\xspace}
\newcommand{\snr}[1][]{signal-to-noise ratio#1 (SNR#1)\renewcommand{\snr}[1][]{SNR##1\xspace}\xspace}

\begin{document}


\title{Magnetic tidal Love numbers clarified}

\newcommand{\napoli}{\affiliation{Sezione INFN Napoli, Complesso Universitario 
di Monte S. Angelo, Via Cinthia, I-80126, 
Napoli, Italy}}
\newcommand{\sapienza}{\affiliation{Dipartimento di Fisica, ``Sapienza'' 
Universit\`a di Roma, Piazzale Aldo Moro 5, 00185, Roma, Italy}}
\newcommand{\infnroma}{\affiliation{Sezione INFN Roma1, Piazzale Aldo Moro 5, 
00185, Roma, Italy}}

\author{Paolo Pani}\email{paolo.pani@roma1.infn.it}\sapienza\infnroma
\author{Leonardo 
Gualtieri}\email{leonardo.gualtieri@roma1.infn.it}\sapienza\infnroma
\author{Tiziano 
Abdelsalhin}\email{tiziano.abdelsalhin@roma1.infn.it}\sapienza\infnroma
\author{Xisco Jim{\'e}nez Forteza}\email{fjimenez@na.infn.it }\napoli\sapienza

\begin{abstract}
In this brief note, we clarify certain aspects related to the magnetic (i.e., 
odd parity or axial) tidal Love numbers of a star in general 
relativity. 
Magnetic tidal deformations of a compact star
had been computed in 2009 independently 
by Damour and Nagar~\cite{Damour:2009vw} and by Binnington and 
Poisson~\cite{Binnington:2009bb}. More recently, Landry and 
Poisson~\cite{Landry:2015cva} showed that the magnetic tidal Love numbers 
depend 
on the assumptions made on the fluid, in particular they are different (and of 
opposite sign) if the fluid is assumed to be in static equilibrium or if it is 
irrotational. We show that the zero-frequency limit of the Regge-Wheeler 
equation forces the fluid to be irrotational. For this reason, the results of 
Damour and Nagar are equivalent to those of Landry and Poisson for an irrotational 
fluid, and are 
expected to be the most appropriate to describe 
realistic configurations. 
\end{abstract}

\maketitle

\section{Introduction} \label{sec:intro}

The deformability of a self-gravitating object immersed in a tidal field is 
measured by the tidal Love numbers~(TLNs)~\cite{PoissonWill}. The theory of 
relativistic TLNs in general relativity has been developed in 
Refs.~\cite{Flanagan:2007ix,Hinderer:2007mb,Binnington:2009bb,Damour:2009vw} 
for nonspinning bodies, and then extended to
rotating bodies 
in~\cite{Poisson:2014gka,Pani:2015hfa,Pani:2015nua,Landry:2015zfa,
Poisson:2016wtv}. This theory has then
been applied to compact binary systems, in order to compute the contribution of 
the tidal deformation to the emitted
gravitational 
waveform~\cite{Vines:2010ca,Vines:2011ud,Damour:2012yf,
Banihashemi:2018xfb,Abdelsalhin:2018reg, 
Landry:2018bil,Jimenez-Forteza:2018buh}. 

For nonspinning objects,\footnote{When the object is 
spinning, angular momentum 
gives rise to spin-tidal coupling and
to a new class of rotational
TLNs~\cite{Pani:2015nua,Landry:2015cva,Landry:2015zfa,Landry:2017piv,
  Gagnon-Bischoff:2017tnz}. In this note we focus on static objects so we shall 
not consider the rotational TLNs.} the TLNs can be separated into 
two classes according to the parity of the perturbation induced by the tidal 
field: induced mass multipole
moments are related to the so-called electric (or even-parity or polar) 
TLNs --~which also exist in Newtonian theory~\cite{PoissonWill}~--, whereas 
induced 
current multipole moments are related to the so-called \emph{magnetic} (or 
odd-parity or axial) TLNs. 
The current multipole moments are induced by an external 
magnetic-type tidal field. Since the latter is not 
a source of the gravitational field in Newton's theory, the magnetic 
TLNs are a genuine prediction of general relativity, which might possibly be 
relevant for very compact objects.

Tidal deformability affects the gravitational-wave phase of a binary 
inspiral at high post-Newtonian order~\cite{Flanagan:2007ix}, with the magnetic 
TLNs giving a small contribution relative to 
the electric ones~\cite{Yagi:2013sva,Jimenez-Forteza:2018buh,Banihashemi:2018xfb}.
Nonetheless, their characterization is important to develop accurate waveform 
models and to compare the post-Newtonian predictions with those of numerical 
simulations~\cite{Dietrich:2017aum,Dietrich:2017feu,Dietrich:2018uni,
Cotesta:2018fcv,Jimenez-Forteza:2018buh}.

There is some confusion in the literature related to the magnetic TLNs.
These were computed independently in 2009 
by Binnington and 
Poisson~\cite{Binnington:2009bb} (hereafter, BP) and by Damour and 
Nagar~\cite{Damour:2009vw} (hereafter, DN) by considering axial perturbations 
of a perfect-fluid star in general relativity (see also~\cite{Favata:2005da} 
for an earlier study by Favata in the context of post-Newtnonian theory). These 
perturbations can be 
reduced to a single second-order master equation; however, it has been 
previously noted that the master equation of BP and that of DN are 
inequivalent~\cite{Pani:2015nua} and give rise to different magnetic TLNs.
Meanwhile, in 2013 Yagi~\cite{Yagi:2013sva} used the result of DN to 
compute the effect of the 
magnetic TLNs in the waveform and to compute some quasi-universal 
relations~\cite{Yagi:2013bca,Yagi:2016bkt} among TLNs of different 
parity and different multipole moments. 
In 2015, Landry and Poisson (hereafter, LP) discovered~\cite{Landry:2015cva} 
that the magnetic TLNs depend on the properties of the
fluid (see also~\cite{Landry:2015snx,Gagnon-Bischoff:2017tnz}). In 
particular, they found that the magnetic TLNs for irrotational
fluids or for static fluids are different and have the opposite sign. 
Consequently, the
quasi-universal relations involving magnetic TLNs also depend on the fluid 
properties 
\cite{Delsate:2015wia,Gagnon-Bischoff:2017tnz,Jimenez-Forteza:2018buh}.

Thus, at the present stage we are left with three different types of magnetic 
TLNs: those computed by DN, those computed by BP, and those computed by LP for 
irrotational fluids.
The scope of this short note is to clarify certain aspects of the magnetic TLNs 
and to unveil the relation between the different magnetic TLNs presented in 
previous work.
As we shall show, the magnetic TLNs computed by DN are actually equivalent (modulo a 
prefactor given in Eq.~\eqref{mapping} below) to those computed by LP for irrotational 
fluids, whereas the magnetic TLNs computed by BP refer to strictly static 
configurations.

\section{Axial perturbations of a perfect-fluid star} \label{sec:setup}
We consider magnetic (i.e., odd parity or axial) perturbations of Einstein's equations in the Regge-Wheeler
gauge~\cite{Regge:1957td}.
In our analysis the perturbations can be time dependent; we
shall analyze the static limit later on. We use geometrical units in which 
$G=c=1$.

We consider a (spherically symmetric) background described by an isotropic 
perfect fluid with stress-energy tensor $T_{\mu\nu}=(\rho+p)u_\mu u_\nu+ p 
g_{\mu\nu}$, where $u^\mu$ is the four-velocity of the fluid, and $p$ and 
$\rho$ are the pressure and the energy density, respectively. 
The background metric, $g_{\mu\nu}^{(0)}dx^\mu dx^\nu = -e^\nu dt^2+e^\lambda 
dr^2+r^2d\Omega^2$, 
satisfies the 
Tolman-Oppenheimer-Volkoff equations,
\begin{equation}
 M'=4\pi r^2\rho,\quad \nu'=2\frac{M+4\pi r^3 p}{r(r-2M)},\quad 
p'=-(p+\rho)\frac{M+4\pi r^3 p}{r(r-2M)}, \label{TOV}
\end{equation}
where a prime denotes a derivative with respect to $r$, and we have defined the 
radial mass function $M(r)$ such that
$e^{-\lambda}=1-2M/r$. In this background, the unperturbed fluid velocity reads $u^\mu=u^\mu_0=\{e^{-\nu/2},0,0,0\}$.

The perturbed metric reads 
$g_{\mu\nu}(t,r,\vartheta,\varphi)= 
g_{\mu\nu}^{(0)}+\delta g_{\mu\nu}^{\rm 
odd}(t,r,\vartheta,\varphi)$, with 
\begin{equation}\label{oddpart}
\delta g_{\mu\nu}^{\rm odd} =\sum_\ell \sum_{m=-l}^l
 \begin{pmatrix}
  0 & 0 & h_0^\ell(t,r) S_\vartheta^{\ell} & h_0^\ell(t,r) S_\varphi^{\ell} \\
  * & 0 & h_1^\ell(t,r) S_\vartheta^{\ell} & h_1^\ell(t,r) S_\varphi^{\ell} \\
  *  & *  & 0 & 0  \\
  * & * & * & 0
 \end{pmatrix}\,,
\end{equation}
where asterisks represent symmetric components, $Y^{\ell}=Y^{\ell}(\vartheta,\varphi)$ are the scalar spherical
harmonics, and 
$(S_\vartheta^{\ell},S_\varphi^{\ell})\equiv\left(-\frac{1} {
\sin\vartheta} Y^{\ell}_{,\varphi}
,\sin\vartheta Y^{\ell}_{,\vartheta}\right)$
are the (odd-parity) vector spherical harmonics.
Since the background is spherically symmetric, the azimuthal number $m$ is degenerate and the perturbation equations
depend only on $\ell$.
Under parity transformations ($\vartheta\rightarrow\pi-\vartheta$, $\varphi\rightarrow\varphi+\pi$), the perturbations
are multiplied by $(-1)^{\ell+1}$ and therefore are called odd-parity or ``axial''; we shall use the two notations
indistinctly.

In the axial sector the metric perturbations are not coupled to pressure and 
density perturbations, but are coupled to
axial fluid perturbations. The only non-vanishing odd-parity fluid perturbation is the axial fluid velocity~(we follow
the notation of Ref.~\cite{Kojima:1992ie} in the nonrotating case):
\begin{equation}
  \delta u^\mu = [4\pi 
e^{-\nu/2}r^2(\rho+p)]^{-1}\,\left(0,0,S_\vartheta^{\ell},\frac{S_\varphi^{\ell}
}{\sin^2\theta} \right)
  U^\ell(t,r)\,,
\end{equation}
such that $u^\mu=u^\mu_0+\delta u^\mu$.
By linearizing Einstein's equations on the background $g_{\mu\nu}^{(0)}$, one 
can obtain a system of three differential equations for the axial sector only
\begin{eqnarray}
 e^{-\nu}\dot h_0-e^{-\lambda} 
h_1'-\frac{1}{r^2}\left(2M-4\pi(\rho-p)r^3\right) h_1&=&0\,, \label{eq1}\\
e^{-\nu}(\dot h_0'-\ddot h_1)-\frac{2 e^{-\nu}}{r}\dot 
h_0-\frac{(l-1)(l+2)}{r^2} h_1&=&0\,,\label{eq2}\\
e^{-\lambda}(h_0''-\dot h_1')-4\pi(\rho+p)r(h_0'-\dot h_1')-\frac{2 
e^{-\lambda}}{r}\dot h_1&& \nonumber\\
-\frac{1}{r^3}(l(l+1)r-4M+8\pi(\rho+p)r^3)h_0-4e^\nu U&=&0\,, \label{eq3}
\end{eqnarray}
where for clarity we omitted the 
multipolar index $\ell$ from the perturbation variables and used a dot to 
denote a time derivative. 

We immediately see that Eq.~\eqref{eq1} can be generically solved for $h_0$ in 
terms of $h_1$, \emph{provided} the perturbations are not strictly 
\emph{static}, in which case $\dot h_0=0$ and Eq.~\eqref{eq1} becomes a 
constraint equation for $h_1$. 

More precisely, Eq.~\eqref{eq1} can be written as
\begin{equation}
 \dot h_0= e^{(\nu-\lambda)/2} (\psi r)'\,, \label{eqh0}
\end{equation}
where $\psi$ is defined such that
\begin{equation}
 h_1=e^{(\lambda-\nu)/2} \psi r\,,
\end{equation}
and we have used the background equations~\eqref{TOV}.
Below, we consider the static and 
time-dependent cases separately.

\subsection{Static axial perturbations}

For strictly \emph{static} perturbations, $\dot h_i=0$ and $U=0$. In this case 
Eq.~\eqref{eq2} yields $h_1=0$, which also satisfies Eq.~\eqref{eq1}. On 
the other hand, Eq.~\eqref{eq3} yields a second-order differential equation
for $h_0$:
\begin{equation}
 e^{-\lambda} h_0''-4\pi 
r(p+\rho) h_0'-\left(\frac{l(l+1)}{r^2}-
\frac{4M}{r^3}+8\pi (p+\rho)\right)h_0=0\,. \label{E1}
\end{equation}
This equation is equivalent to that obtained by BP (cf. 
Eq.~(4.29) in Ref.~\cite{Binnington:2009bb}) which indeed studied the axial 
perturbations of a strictly static fluid.

Although Ref.~\cite{Binnington:2009bb} reported that Eq.~\eqref{E1} is also 
equivalent to 
Eq.~(31) in DN~\cite{Damour:2009vw}, this is actually not the case, as 
already noticed in Ref.~\cite{Pani:2015nua}. We shall elucidate the reason for 
this discrepancy in the next section.

\subsection{Time-dependent axial perturbations}
Let us consider the
Fourier transform of the perturbations, i.e.  $h_i(t,r)=\int dt\, h_i(\omega,r)e^{-i\omega t}$, with a slight abuse of
notation. In this case Eq.~\eqref{eqh0} can be solved for $h_0$ in terms of $h_1$ and its derivative:
\begin{equation}
 h_0(\omega,r)= i\frac{e^{(\nu-\lambda)/2}}{\omega} (\psi r)'\,. \label{eqh0b}
\end{equation}
Notice that the above equation does not have a well-defined limit 
as $\omega\to0$. 
Inserting Eq.~\eqref{eqh0b} into Eq.~\eqref{eq2} yields 
\begin{eqnarray}
&&e^{(\nu-\lambda)/2}(e^{(\nu-\lambda)/2}
\psi')' \nonumber\\
&+&\left[\omega^2-e^\nu\left(\frac{l(l+1)}{r^2}-\frac{6M}{r^3}
+4\pi(\rho-p)\right)\right]\psi=0\,,\label{E2}
\end{eqnarray}
which is the standard Regge-Wheeler equations for axial perturbations inside the star (see
e.g. Ref.~\cite{Kojima:1992ie}). In the limit $\omega\to0$ this equation 
coincides with Eq.~(31) in
DN~\cite{Damour:2009vw}.

We shall now show that the $\omega\to0$ limit of Eq.~\eqref{E2} is inequivalent 
to Eq.~\eqref{E1}.
The underlying reason for this fact can be traced back to the perturbation 
of the fluid velocity, which for $\omega\neq0$ is (see, 
e.g.,~\cite{Kojima:1992ie})
\begin{equation}
 U=-4\pi(\rho+p)e^{-\nu}h_0\,.\label{eqU}
\end{equation}
The above equation can be obtained by an appropriate combination of the 
components of Einstein's equations or, more
directly, by the axial component of the stress-energy tensor conservation.
Therefore, even when $\omega\to0$ the fluid velocity is nonvanishing and the 
configuration is not strictly static.
By replacing Eq.~\eqref{eqU} into Eq.~\eqref{eq3}, it is straightforward to 
obtain an equation for $h_0$ which, in the limit $\omega\to0$, reads
\begin{equation}
 e^{-\lambda} h_0''-4\pi 
r(p+\rho) h_0'-\left(\frac{l(l+1)}{r^2}-
\frac{4M}{r^3}-8\pi (p+\rho)\right)h_0=0\,. \label{E1b}
\end{equation}
This equation coincides with Eq.~(5.6) in LP for an \emph{irrotational} 
fluid ($\lambda=1$ in
their notation). As noticed in LP, Eq.~\eqref{E1b} is actually very similar 
to Eq.~\eqref{E1} for the static case, the only difference being the
opposite sign in front of the $(\rho+p)$ term.

One can easily check that the fluid in this configuration is 
\emph{irrotational}, i.e. the vorticity vector
$\omega^\alpha=\frac{1}{2}\epsilon^{\alpha\beta\mu\nu}u_{\beta;\mu}u_\nu$ identically
vanishes~\cite{synge1938relativistic}. This corresponds to the configuration 
studied by LP~\cite{Landry:2015cva}. In our case this condition is enforced by 
Eq.~\eqref{eqU}, while in the static case $U=0$.

The fact that Eq.~\eqref{E1} and Eq.~\eqref{E1b} are inequivalent shows that 
the limit $\omega\to0$ of the axial sector is discontinuous, i.e. in this  
limit the Regge-Wheeler equation is not equivalent to Eq.~\eqref{E1} which 
describes the static case, 
$\omega=0=U$. The latter is an isolated point in the space of the solutions.

\section{Discussion}
\label{sec:conclusions}

In summary, we showed that the equation describing the magnetic TLNs computed by DN coincide 
with that computed by LP 
for an irrotational fluid. This is due to the zero-frequency limit of the 
Regge-Wheeler equation, which forces the fluid to be irrotational rather than 
static. This fact also explains why the master equations computed by DN and by 
BP are inequivalent, because in the former case the fluid is irrotational, 
whereas in the latter case the fluid is static.
To the best of our knowledge, this connection was not pointed out in the past. 

In particular, the relation between the magnetic TLNs computed by DN 
(denoted as $j_{\ell}$) and those computed by LP (denoted as $\tilde k_{\ell}^{\rm mag}$) for an 
irrotational fluid is (see also Eq.~(6) in Ref.~\cite{Banihashemi:2018xfb} for the 
$\ell=2$ case)
\begin{equation}
  j_\ell =\frac{4(\ell+2)(\ell+1)}{\ell(\ell-1)}\left(\frac{{\cal M}}{{\cal R}}\right) \tilde k_\ell^{\rm 
 mag}\,.\label{mapping}
\end{equation}
Note that the two definitions differ by a factor of the 
compactness, ${\cal M}/{\cal R}$, where ${\cal M}=M({\cal R})$ and ${\cal R}$ are the 
stellar mass and radius, respectively.

Yagi~\cite{Yagi:2013sva} used the master equation derived
by DN so he actually computed the magnetic TLNs $j_\ell$ which, as we have just shown, 
correspond to the case of an irrotational fluid. 
In particular, the static and irrotational magnetic TLNs satisfy two different 
approximately-universal relations, as discussed in 
Refs.~\cite{Gagnon-Bischoff:2017tnz,Jimenez-Forteza:2018buh}, where some fits 
for such relations are 
provided in both cases.

Finally, since the irrotational case is 
obtained as the zero-frequency limit of the Regge-Wheeler equation, we 
consider it to be more physical, which is also on the line of recent numerical 
relativity simulations of binary neutron star mergers 
\cite{Bonazzola:1998yq,Bernuzzi:2011aq,Tichy:2016vmv,Haas:2016cop}, and 
therefore we expect it should describe more 
accurately relevant astrophysical 
configurations~\cite{Shapiro:1996up,Favata:2005da,Landry:2015cva}

It is also worth mentioning that the magnetic TLNs of static and of 
irrotational fluids are similar in absolute values (and of opposite sign). This 
implies that in both cases their contribution to the waveform is very 
small, and might be possibly be relevant only for third-generation gravitational-wave 
detectors, as recently analyzed in detail~\cite{Jimenez-Forteza:2018buh}.

\vspace{-\baselineskip}

\section*{Acknowledgments}
We thank Philippe Landry for interesting discussion and for pointing out  
Eq.~\eqref{mapping} in a private comment on our draft.
P.P. acknowledges financial support provided under the European Union's H2020 
ERC, Starting Grant agreement no.~DarkGRA--757480.
The authors would like to acknowledge networking support by the COST Action 
CA16104 and by the European Union's Horizon 2020 
research and innovation programme under the
Marie Sklodowska-Curie grant agreement No 690904.
We acknowledge support from the Amaldi Research Center funded by the
MIUR program "Dipartimento di Eccellenza" (CUP: B81I18001170001).

\bibliography{biblio.bib}

\end{document}